# Time and Gravity

*Robert Sadykov*
Department of Relativity Theory and Gravity, Kazan University, Russia
*robertsadykov@gmail.com*

### Abstract

Several physical concepts, including the concept of time, are clarified herein by taking into account existing experimental data. In addition, the missing links among these physical concepts are established. This allows us to take another step towards understanding the physical nature of time.

### Keywords

time – space – gravitation – relativity

### Introduction

We can approach the understanding of time in the context of physics if we find a direct relationship of time with other physical concepts such as space, motion, energy, momentum, inertia and gravitation. This list can be expanded in the future, and a more precise definition of relations between all elements of the list may also be required. The problem, however, is that not all of these concepts are unambiguous. For example, energy is presented in potential and kinetic forms. The relation of kinetic energy with space, motion, inertia and momentum is more or less obvious; but potential energy, including the internal energy of particles, remains a mystery. Of course, the biggest mystery is gravitation. It is with its definition, based on the experimental data available today, that we shall begin. As a first step, we temporarily abstract from the Newtonian theory of gravity and the general theory of relativity. In particular, this means that we shall not refer to the classical gravitational forces. In the process of forming a new concept of gravitation, we can be based upon the special theory of relativity. However, the area of gravitation may be wider than previously thought and may fully include the area that today refers to special relativity. Therefore, for greater objectivity, we also abstract from special relativity. As a next step, we appeal to the features of the propagation of light in three different locations:

1) The constant speed of light in interstellar space—this is confirmed by observing binary stars, astrophysical jets and accelerated fragments of supernova; observations show that the speed of light is not summed up with the speed of the light source and remains constant relative to the system of surrounding stars;
2) Some decrease in the speed of light in the gravitational fields of planets, stars and star clusters. This is confirmed, in particular, by the delay of light in the gravitational field of the Sun;
3) The constant local speed of light on Earth—this is confirmed by measuring the speed of light emitted by accelerated particles.

All three locations include the surrounding masses. Therefore, we can assume that gravitation forms not only the observed variable speed of light (item 2) but also the observed constant speed of light (items 1 and 3). In other words, the constancy of the speed of light can be a special case of the action of gravitation. Before we formulate this assumption as a postulate, we shall agree on terminology.

## 1 Terminology

The inertial properties of a proton or an electron can be described by two classical definitions of inertial mass. The first definition is related to force and acceleration: $m = F/a$. The second definition of mass describes mass through momentum and speed: $m = p/v$. Both definitions originate in Newtonian mechanics. The first definition cannot be applied to a quantum of electromagnetic radiation—a photon—due to the impossibility of force action on the photon. For this reason, as applied to the photon, the choice focuses on the second definition of mass, where we find a nonzero value of this physical quantity. For this type of mass, a new physical term is introduced—kinetic mass. This term takes into account the fact that the photon exists only in a state of motion. Alternatively, this mass can be called an effective mass; but the term kinetic mass better reflects the relationship of this mass with motion. It should be noted that photon momentum used to be expressed through its energy $p = E/c$; conversely, photon energy was expressed through its momentum $E = pc$. This is equivalent to describing the momentum through … the momentum. Now, with the appearance of a new physical quantity, photon momentum is represented by fairly simple elements—speed and kinetic mass: $p = mc$. Photon energy is also described in terms of kinetic mass and speed:

$E = mc^2$. If someone still thinks that a photon is a particle with zero rest mass, then, first of all, he needs to determine what definition of the mass he uses. The first definition of inertial mass works flawlessly in Newtonian mechanics; but it is absolutely incompatible with the observed properties of the photon. Within the first definition, zero mass means no resistance to acceleration, but acceleration itself is allowed. As already noted, a photon cannot be accelerated with the participation of forces. In addition, the photon does not exist in a stationary state and any reference to this state in determining the characteristics of a photon goes beyond physics. The second definition of inertial mass considered, on the contrary, is fully compatible with the photon and completely excludes the zero value of the photon mass. So, the kinetic mass of the photon, together with the speed, form the momentum and kinetic energy of the photon. The massless photon, like its predecessor, the weightless ether, remains in the past.

## 2 Postulates

Based on the behavior of light in gravitational fields, we postulate that gravitation sets the relevant speed of light relative to the source of gravitation. Setting the relevant speed of the photon is realized together with the change in the kinetic mass of the photon (see Section 1). This effect will be called the gravidynamic effect. Consider the basic implementation of this effect on the following physical model. The photon moves from point $A$ (Figure 1), located far enough from the central mass $M$ to point $B$, located on the surface of this mass.

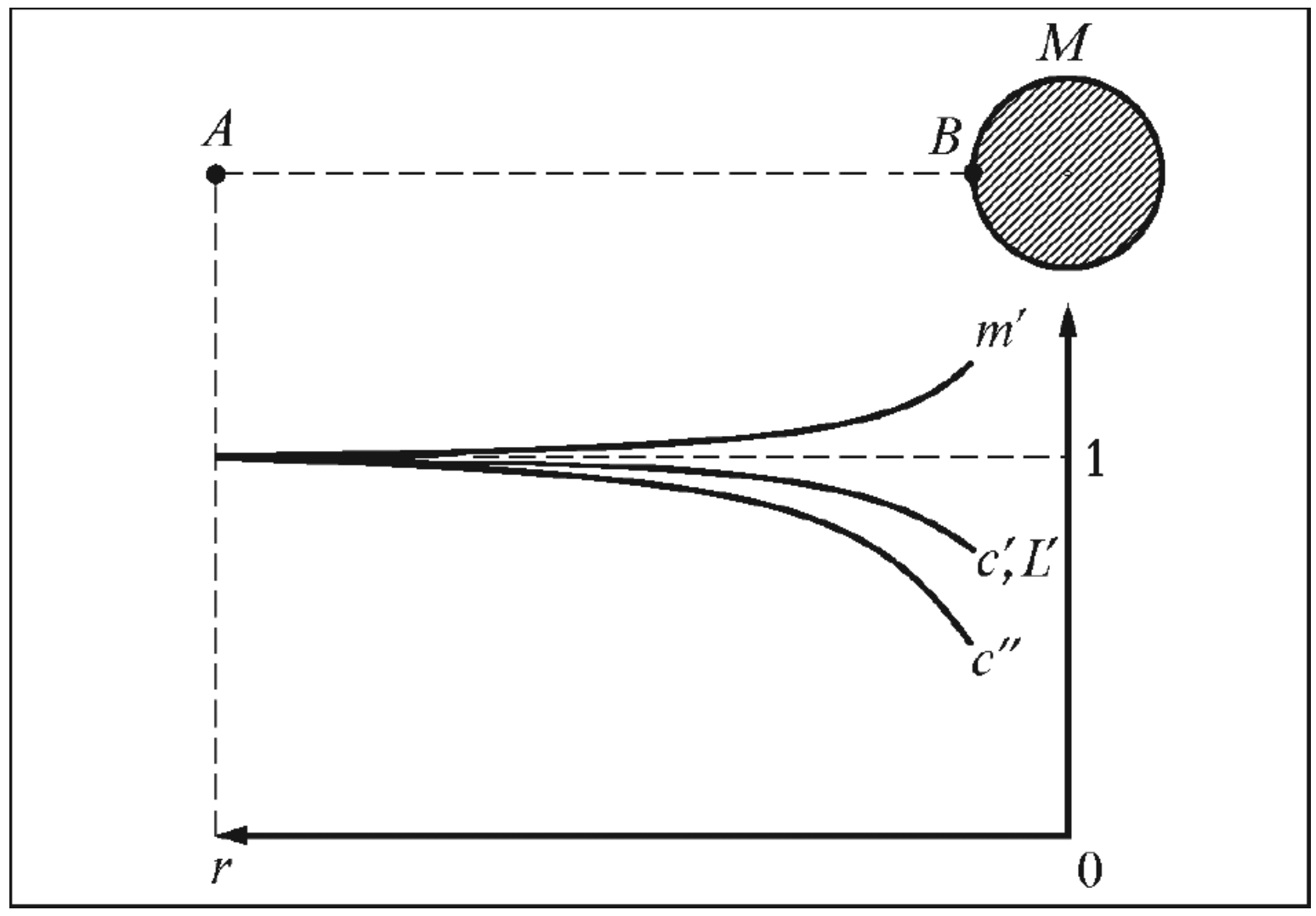

FIGURE 1 Gravitational field of the central mass

During the motion of the photon to point $B$, the gravidynamic effect increases the kinetic mass of the photon:

$$m' = m + \Delta m = m\,(1 + GM/rc^2) \qquad (1)$$

where $r$ is the distance from the photon to the center of the mass $M$, and $G$ is the gravitational constant. In this abstract physical model, factors that could change the value of the photon's momentum are absent. The gravidynamic effect is the single effect that characterizes gravitation at this stage. From the point of view of a remote observer situated at point $A$, an increase in the kinetic mass of a photon in full accordance with the law of conservation of momentum is accompanied by a proportional decrease in its speed:

$$c' = mc/m' = c/(1 + GM/rc^2) \qquad (2)$$

The maximum value of the kinetic mass of the photon $m'$ and its minimum speed $c'$ are reached at point $B$. After reflection at point $B$, the photon moves in the direction of point $A$. In this period, the gravidynamic effect reduces the kinetic mass of the photon and thereby increases its speed. Finally, at point $A$, the kinetic mass and speed of the photon restore the initial values $m$ and $c$. The remote observer can indirectly observe the change in the photon speed as some delay in the photon's return. Looking ahead, we note that the local observer situated at point $B$ has a slightly different point of view on what is happening.

It is a fact that the gravidynamic effect explains only half the delay of the electromagnetic signal in the gravitational field of the Sun observed during the radiolocation of planets. Therefore, in addition to the gravidynamic effect, we postulate the gravicontraction effect, which consists in the contraction of local space:

$$L' = L\,/\,(1 + GM\,/\,rc^2) \qquad (3)$$

where $L'$ is the actual local length. From the point of view of the remote observer situated at point $A$, each meter from point $A$ to point $B$ is reduced to varying degrees by this effect; but at the same time the total number of meters increases. This leads to an elongation of the trajectory of the photon. Elongation of the trajectory, in turn, leads to some delay in the return of the photon to point $A$, and this is added to the delay caused by the gravidynamic effect. Thus, as a result of the combined action of the gravicontraction and gravidynamic effects, the effective speed of the photon becomes equal to

$$c'' = c / (1 + 2GM / rc^2 + G^2M^2 / r^2c^4) \quad (4)$$

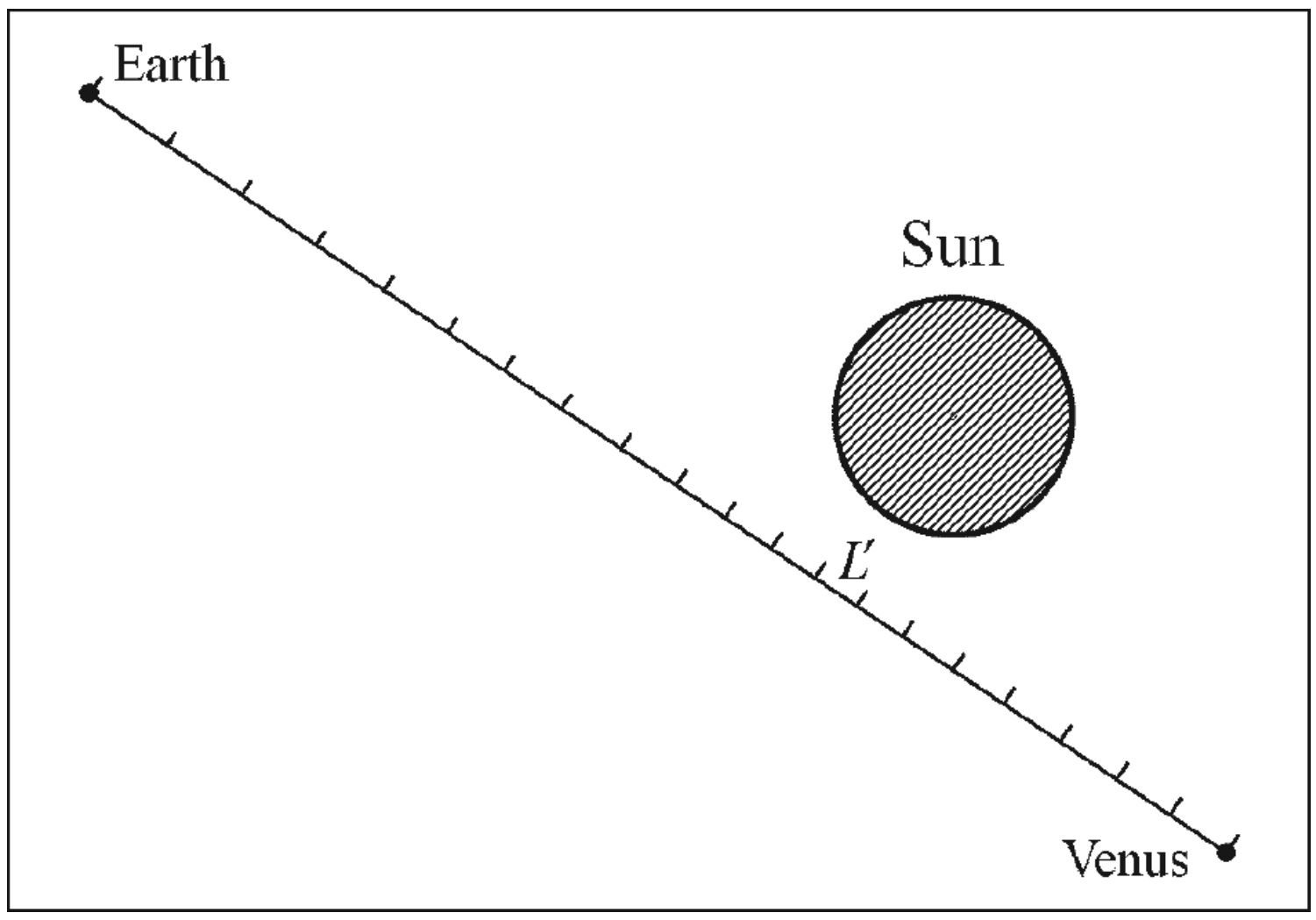

FIGURE 2 Gravitational time delay of an electromagnetic signal at a radiolocation of Venus

It is this resultant speed that forms the observed delay of the electromagnetic signal in the gravitational field of the Sun (Figure 2).

It should be noted that in the neighborhood of mass $M$ (Figure 1), not only space, but also any objects located in this space are reduced in three spatial dimensions in proportion to the contraction of the local length $L'$.

We have, then, discovered two hidden gravitational effects and, for further detailed research, we accepted them as postulates. Can the consequences obtained from these postulates explain the accelerated fall of the test particle or, for example, the deflection of light in a gravitational field?

## 3 Gravitational Refraction

Because of the dependence of the effective speed of light $c''$ on the gravidynamic and gravicontraction effects (see Section 2), the gravitational field of central mass $M$ (Figure 1) acts as a refracting medium with a variable index of refraction:

$$n = 1 + 2GM / rc^2 + G^2M^2 / r^2c^4 \quad (5)$$

The angle of refraction of light in weak gravitational fields, for example, in the gravitational field of the Sun (Figure 3), is equal to

$$\alpha = 4GM / rc^2 \tag{6}$$

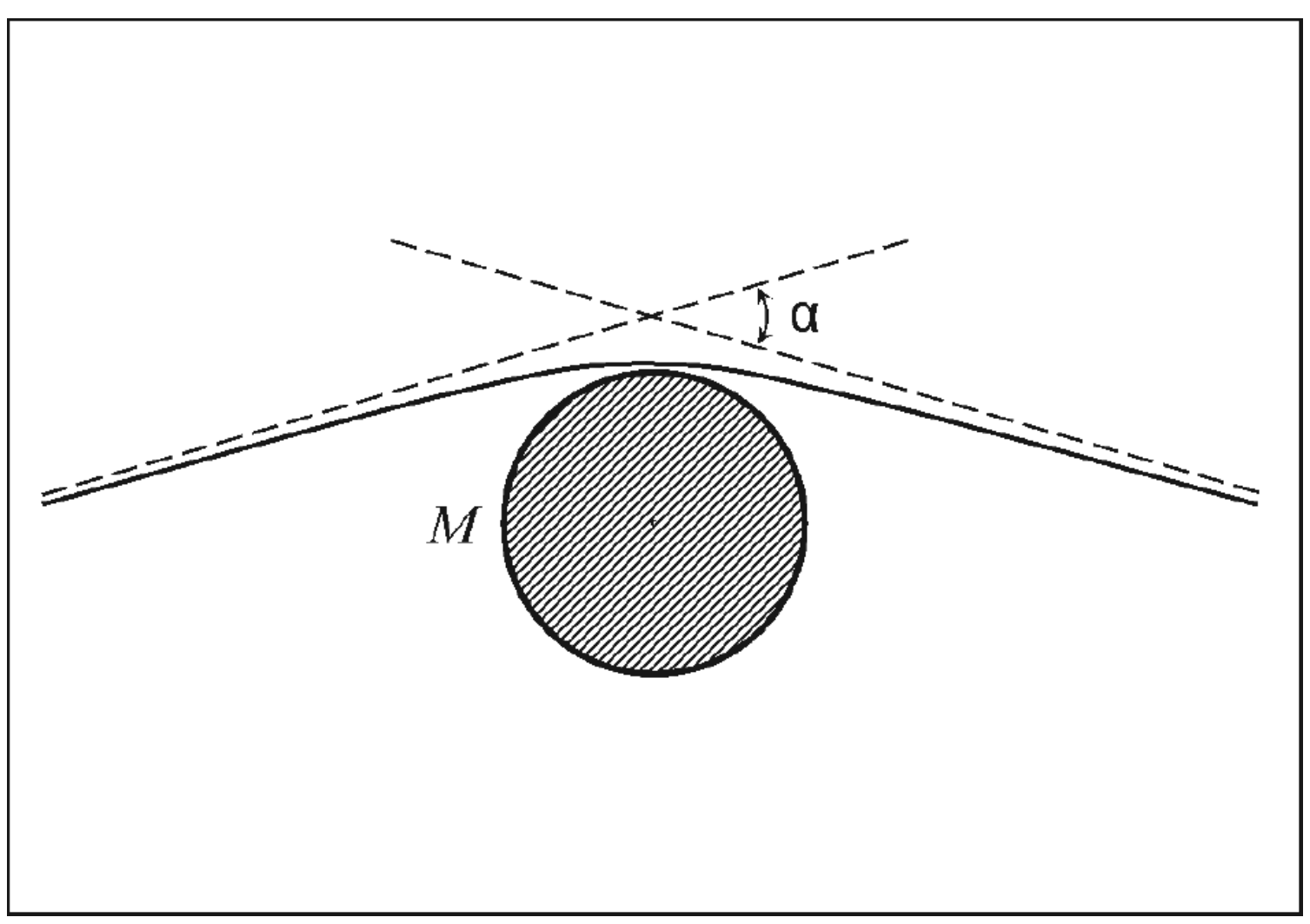

FIGURE 3 Gravitational refraction of light

Where $r$ is the minimal distance of the light ray from the center of mass $M$.

The gravitational refraction of light is therefore a consequence of the gravidynamic and gravicontraction effects.

## 4 Gravitational Acceleration

Let a large number of photons move inside the reflecting sphere forming a spherical photon cluster. The kinetic energy of photons is the internal energy of this cluster:

$$E_{(S)} = c^2 \sum m \tag{7}$$

The kinetic mass of photons forms the classical inertial mass of this cluster:

$$m_{(S)} = \sum m \tag{8}$$

Such a photon cluster can also be considered as a three-dimensional light clock. Thus, the photon cluster can simulate the energy, inertial and temporal properties of an electron and other particles with a classical inertial mass. Let the photon cluster be located in the gravitational field of the central mass $M$ (Figure 4).

Because of the combined action of the gravidynamic and gravicontraction effects, the gravitational field acts as the refracting medium with the variable index of refraction $n$ (see Section 3). Gravitational refraction of the trajectory of the photons inside the cluster leads to acceleration of the cluster in the direction of mass $M$. So, the gravitational acceleration of the photon cluster or any test particle with a classical inertial mass is a consequence of the gravidynamic and gravicontraction effects.

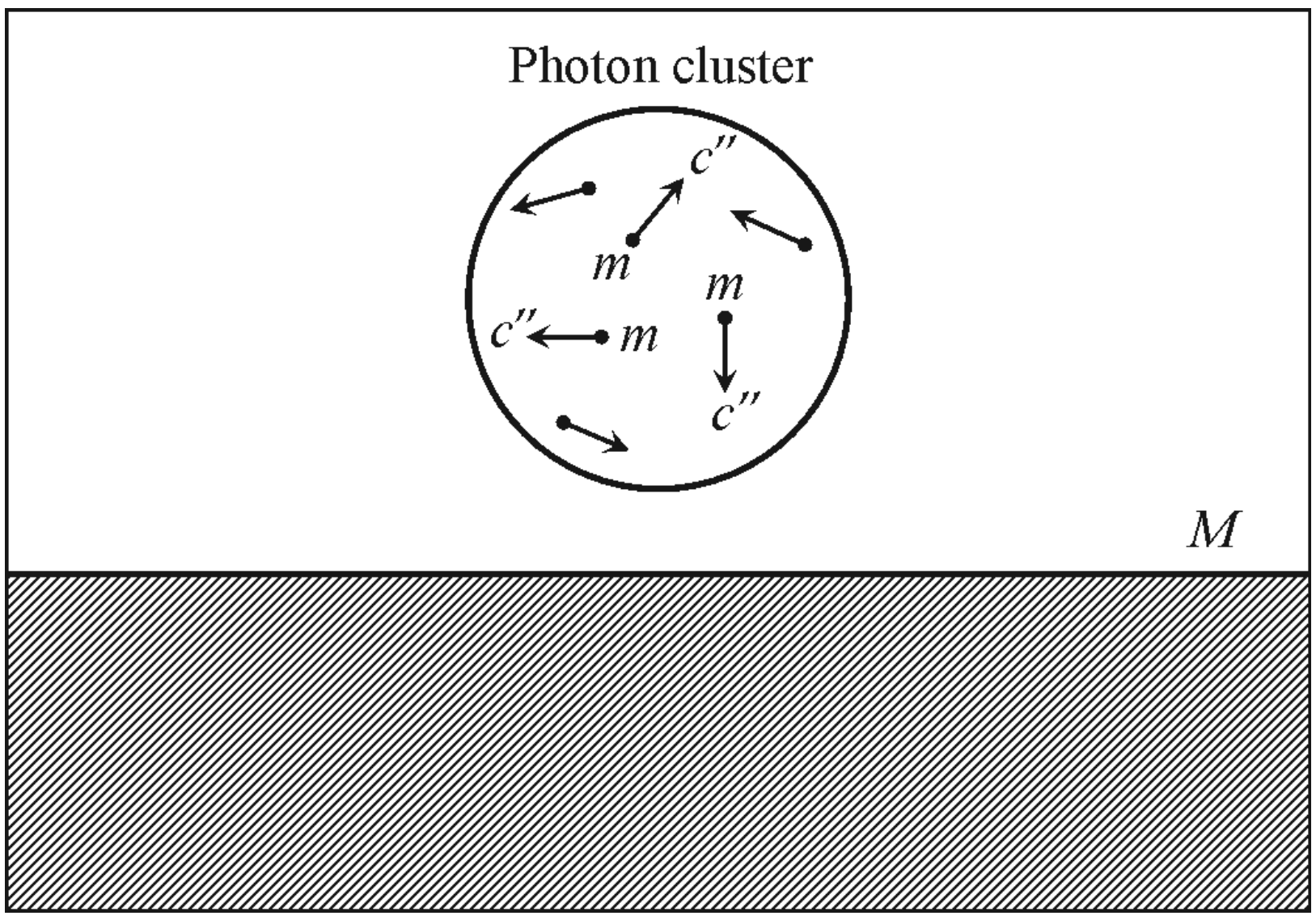

FIGURE 4 Photon cluster in the gravitational field

## 5 Features of Gravitational Acceleration

The gravitational refraction of photons inside the photon cluster (Figure 4) has its own specifics. Refraction is completely absent for photons moving in the gravitational field in the vertical direction. Photons moving in the horizontal direction have the maximum refraction. The refraction of many other photons has various intermediate values. These factors cause a significant dependence of the gravitational acceleration of the photon cluster on the speed and direction of motion of the cluster relative to the mass $M$. Let us consider three characteristic cases:

1) If the photon cluster moves at a relatively low speed, then the gravitational acceleration of the cluster fully coincides with the standard Newtonian value;
2) If the photon cluster moves in the vertical direction at a high speed, then the gravitational acceleration of the cluster is less than the standard value;
3) If the photon cluster moves in the horizontal direction at a speed close to the speed of light (Figure 5), then the gravitational refraction for most of the photons reaches the maximum possible value. Therefore, the gravitational acceleration of the cluster is twice the standard Newtonian value.

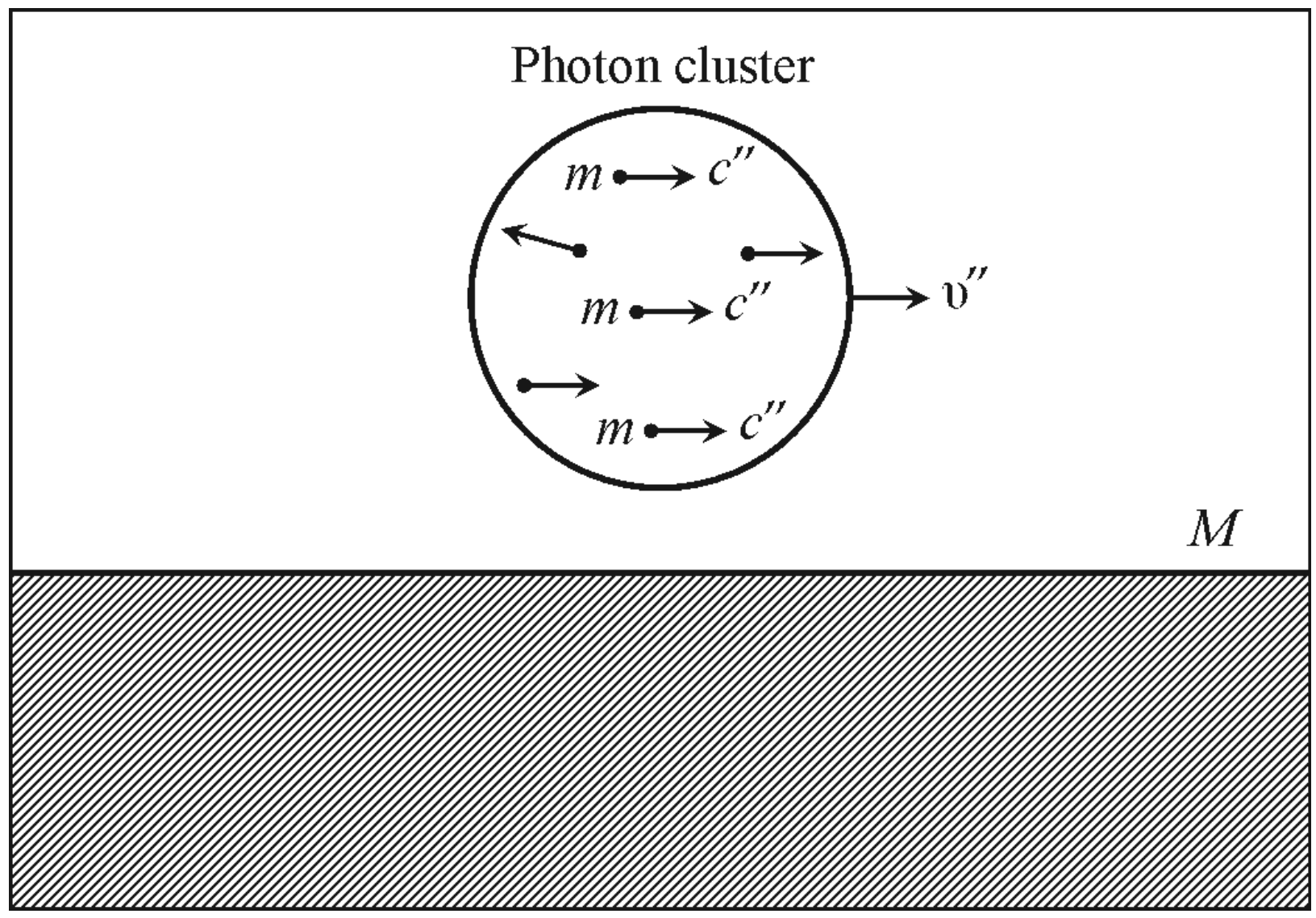

FIGURE 5 Motion of the photon cluster in a gravitational field

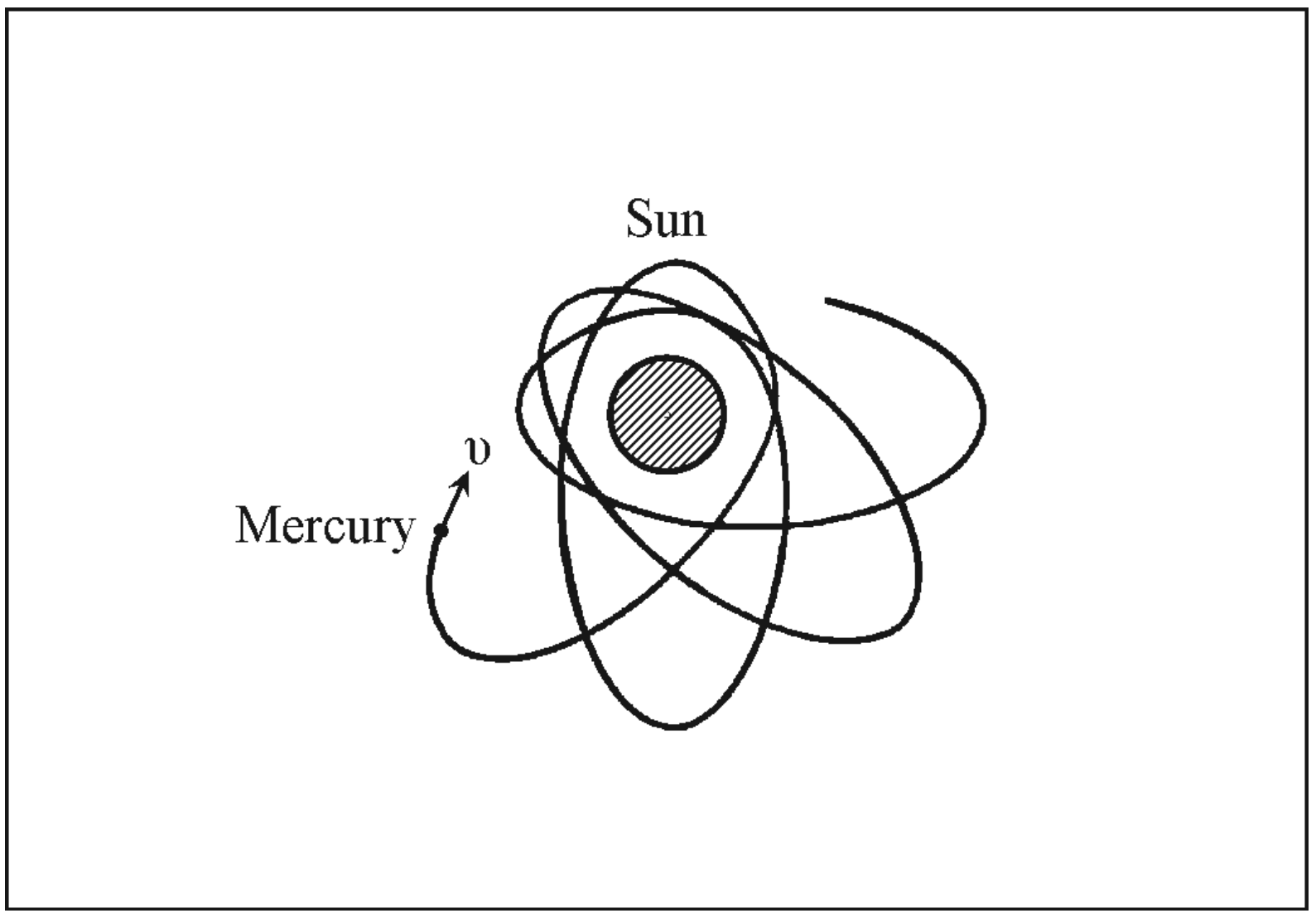

FIGURE 6 The perihelion precession of Mercury

This last case can be observed as a double angle of the deviation of the test mass moving at the sub-light speed in the gravitational field of the Sun. In the case of the comparatively low speed of Mercury relative to the Sun, a small increase in the gravitational acceleration in comparison with the standard value still exists, which is observed as an anomalous shift of the Mercury perihelion (Figure 6).

## 6 Gravitational Time Dilation

The idea of combining space and time into a single four-dimensional space-time continuum is a good idea from the point of view of mathematics, but raises some questions from the point of view of physics. For example, is there time in space in the absence of particles and fields, and if so, is there time at every point in space? Due to the lack of unambiguous answers to these questions, we will stick to another promising concept of time, where one-dimensional time is a simple consequence of a change in the state of a physical system. Since the state change is usually described in terms of time, the question arises: what is primary, a state change or time? The answer is unequivocal: a change of state is primary in relation to time. In a physical system where any state change is missing, time simply stops.

The proper time of the photon cluster (see Section 4) is formed by a change in state of this cluster, which is caused by the movement of all photons forming the cluster. The transfer of the photon cluster from point $A$ to point $B$ (Figure 1) is accompanied by a decrease in the speed of photons $c'$ due to the gravidynamic effect (see Section 2). This slows down the state change of the cluster, which in turn leads to a dilation of the proper time of the cluster:

$$T' = Tc/c' = T(1 + GM/rc^2) \tag{9}$$

The same thing happens with ordinary light clocks. The retarding action of the gravidynamic effect at point $B$ is not limited to the light clock and the photon cluster, since the speed of light is a characteristic speed not only for the propagation of light but also for other physical processes. In other words, a decrease in the local speed of light causes a dilation of the local time. It should be noted that for a remote observer situated at point $A$, the photon speed at point $B$ is $c'$ if he takes into account the decrease in the local length $L'$ at this point and $c''$ if the change in the local length is ignored. For a local observer situated at point $B$, the local speed of the photon is equal to the typical speed of light $c$. The reason for this is quite simple and consists in the following: per unit of local time dilated by the gravidynamic effect, the photon slowed down by this effect covers the local distance $3 \times 10^8$ m typical for light.

## 7 Location of Gravitational Energy

The gravitational refraction, like any other refraction, does not perform mechanical work. But the gravitational refraction of photons inside the photon cluster (Figure 4) causes an accelerated shift of the center of inertia of the cluster towards the central mass $M$. This means that the internal energy of the photon cluster is converted into external kinetic energy. This converting leads to a decrease in the internal energy of the cluster:

$$E'_{(S)} = E_{(S)} / (1 + GM / rc^2) \tag{10}$$

The total energy remains constant. So, the gravitational energy that is used for the acceleration of the test particle to the central mass is not located in the gravitational field and certainly not in the central mass, but is concentrated in the test particle itself. The gravitational field of the central mass only initiates the conversion of the internal energy of the test particle $mc^2$ into external kinetic energy, but does not change the total energy of this particle. On the contrary, if the test particle moves by inertia from point $B$ in the direction of point $A$, then the reverse process takes place: the kinetic energy of the test particle is converted into the internal energy of this particle. This conversion is accompanied by a decrease in the kinetic energy of the particle. Thus, the test particle in the process of gravitational acceleration does not remain unchanged and this essentially distinguishes the gravitational acceleration from the ordinary acceleration involving forces. It should be noted that the decrease in the internal energy of the test particle at point $B$, as well as the dilation of the proper time of this particle (see Section 6) are not observed locally due to the strictly identical action of the gravitational field on all local particles, as well as on the local observer.

## 8 Gravitational Redshift

We assumed that the photon gradually loses its energy when moving in a gravitational field vertically upwards and, on the contrary, receives energy when moving vertically downwards. However, the reality is somewhat different. First of all, the gravitational field is not a classical potential field (see Sections 2 and 7). The gravidynamic effect dilates the proper time of the light source $T'$ (see Section 6) at point $B$ (Figure 1), which leads to a decrease in the frequency of the emitted light. This decrease in frequency is observed at point $A$ as a gravitational redshift. In other words, the frequency and energy of an emitted photon are reduced at point $B$ from the very beginning and do not change during the movement of a photon from point $B$ to point $A$.

## 9 Constancy of the Local Speed of Light

Let the test particle move at a speed $\upsilon'$ in the vicinity of point $B$ (Figure 1). The remote observer is situated at point $A$. By condition, the speed $\upsilon'$ is much less than the relevant speed of light $c'$ at point $B$ (see Section 2). The photon is emitted by the test particle in the direction of particle motion. According to the classical, non-relativistic law of addition of speeds, the photon speed is added to the speed of the emitting particle, but the gravidynamic effect fixes the photon speed relative to the central mass $M$ in the value $c'$ by increasing the kinetic mass of the photon:

$$m'' = m'/(1 - \upsilon'/c') \tag{11}$$

In the case of emission of the photon in the opposite direction, the gravidynamic effect decreases its kinetic mass:

$$m'' = m'/(1 + \upsilon'/c') \tag{12}$$

and thereby also fixes the photon speed in the value $c'$. If a photon is emitted in other directions, then the kinetic mass of the photon takes on various intermediate values:

$$m'' = m'/(1 - (\upsilon'/c') \cos \theta) \tag{13}$$

where $\theta$ is the angle between the motion direction of the emitting particle and the emitting direction. In all cases, the photon speed remains fixed relative to the central mass. So, any attempt to change the relevant speed of the photon $c'$ by changing the speed of the light source or the direction of the emission is compensated by a corresponding change in the kinetic mass of the photon. As noted earlier, for the local observer situated at point $B$, the local speed of the photon has a typical value $c$ (see Section 6).

## 10 Dependence of Inertial Mass on the Action of Gravitation

Let the photon cluster be accelerated in the gravitational field of the central mass $M$ (Figure 5) to speed $\upsilon'$. Despite the acceleration of the cluster, the gravidynamic effect fixes the relevant speed of each photon $c'$ relative to the central mass by increasing or decreasing the kinetic mass of photons (see Section 9), depending on the direction of their movement after each reflection. In the process of acceleration of the photon cluster, the total kinetic mass of the photons increases and this leads to an increase in the inertial mass of the cluster:

$$m''_{(S)} = m'_{(S)} / \sqrt{1 - \upsilon'^2 / c'^2} \tag{14}$$

The increase of inertial mass of the photon cluster or any test particle accelerated relative to the central mass is therefore a consequence of the gravidynamic effect.

The gravidynamic effect decreases the relevant speed of the photon at point $B$ (Figure 1) to the value $c'$ and, as a result, dilates the proper time of the light clock $T'$ located at this point (see Section 6). If the light clock moves at a constant speed $\upsilon'$ relative to the central mass $M$, then the gravidynamic effect causes an additional dilation of the proper time of this clock:

$$T'' = T' / \sqrt{1 - \upsilon'^2 / c'^2} \qquad (15)$$

This is due to the fact that the gravidynamic effect fixes the relevant speed of the photon $c'$ relative to the central mass and thereby reduces the speed of the photon relative to the light clock (Figure 7).

If a moving light clock is replaced by a photon cluster, then the gravidynamic effect fixes the relevant speed of all photons $c'$ relative to the central mass, regardless of the direction of their movement (see Section 9). As a consequence, the average speed of photons inside the cluster decreases, which leads to the dilation of the proper time of the photon cluster. The flawless work of the gravidynamic effect initially violates the symmetry in the speeds of time flows and thereby excludes the appearance of a contradiction known as the clock paradox. The time dilation, then, is always a consequence of the gravidynamic effect and is realized in two slightly different forms. The first form is observed when the test particle is stationary in the gravitational field of the central mass. The second, additional form occurs when the test particle begins to move in the gravitational field of the central mass.

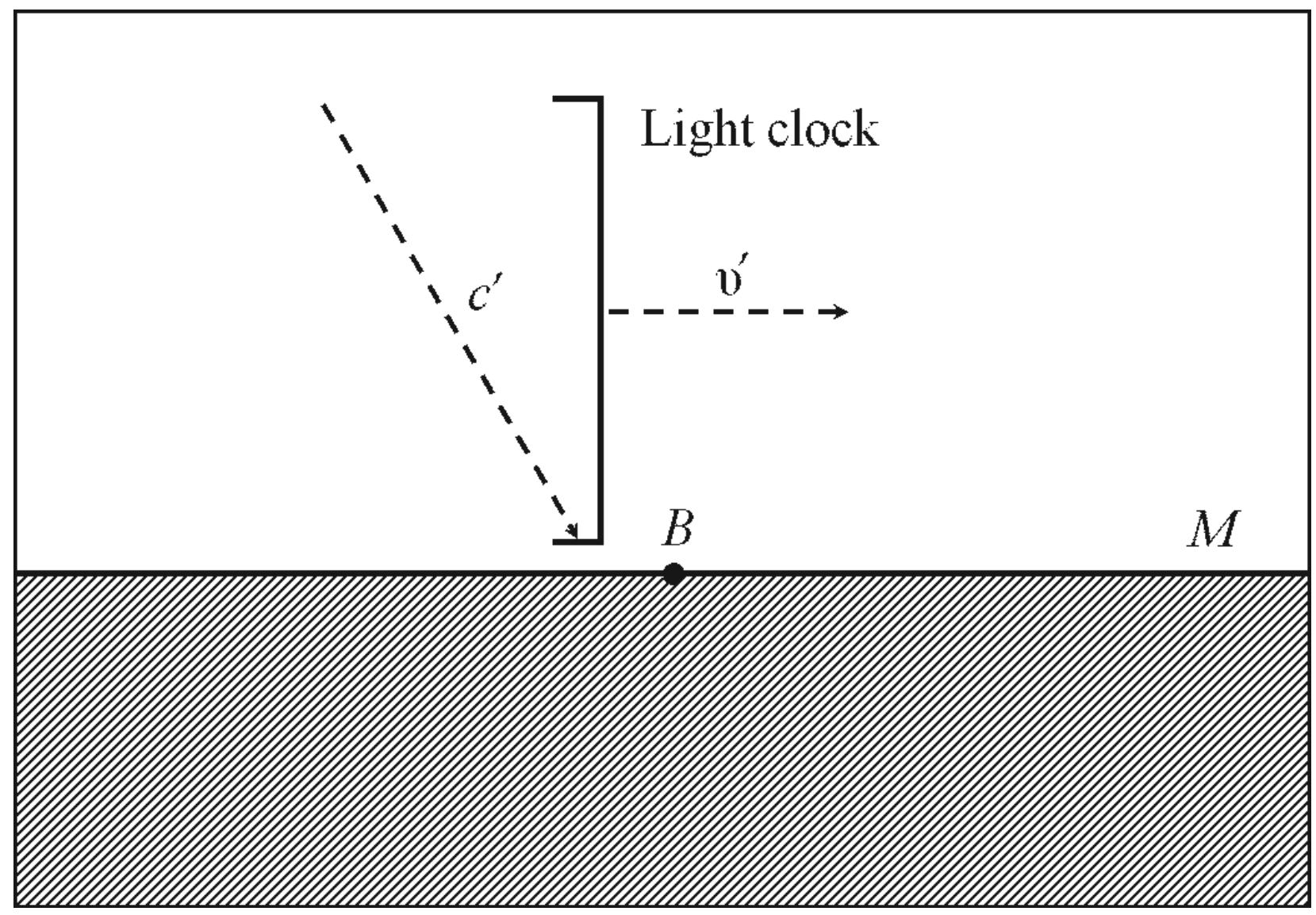

FIGURE 7 Moving light clock

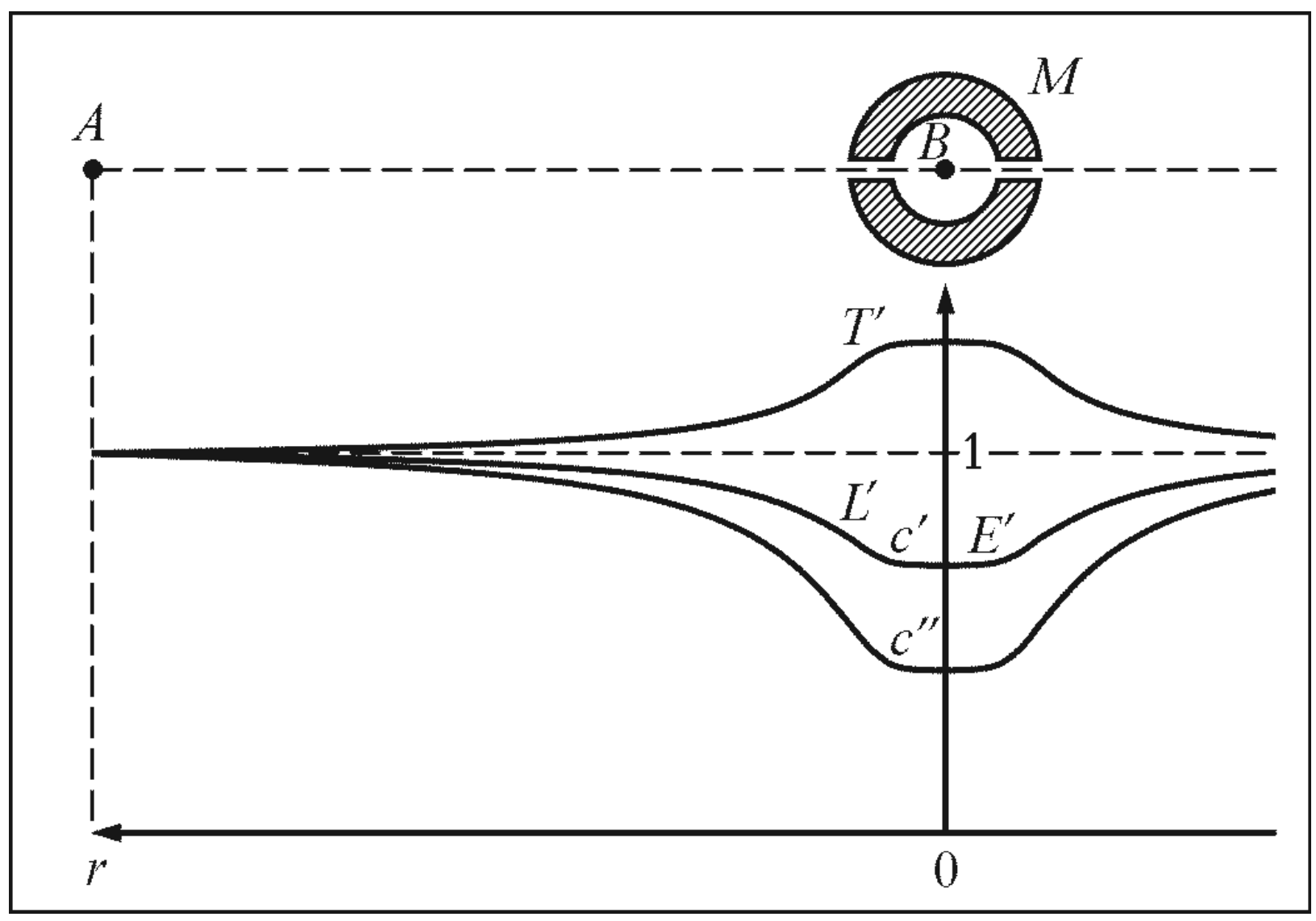

FIGURE 8 Speed of light in a gravitational field of spherically symmetric mass

## 12 Constancy of the Speed of Light in Interstellar Space

Let the photon move from remote point *A* to point *B*, which is located in the central hollow area of the spherically symmetric mass *M* (Figure 8). In the process, the gravidynamic effect gradually decreases the relevant speed of the photon $c'$ (see Section 2). In addition to this, the gravicontraction effect lengthens the photon trajectory and thereby decreases the effective speed of the photon $c''$ (see Section 2). These speeds reach a minimum in the empty central area (see Diagram in Figure 8). Two secondary gravitational effects—gravitational refraction and gravitational acceleration (see Sections 3 and 4) are completely absent in the central area due to the constant effective speed of light $c''$ in this area. However, other gravitational effects continue to act. For example, the relevant speed of light $c'$ in the central area remains constant relative to mass *M* even if the light is emitted by a fast moving particle (see Section 9). This speed is *c* for a local observer stationary relative to mass *M* (see Section 6). The internal energy of a test particle, for example an electron, decreases in this area to the value $E'$ (see Section 7). Parallel to this, the proper time of the electron dilates to value $T'$ (see Section 6) and additionally dilates to value $T''$ in the case of electron motion relative to mass *M* (see Section 11). In addition, the motion of the electron relative to mass *M* causes an increase in the inertial mass of the electron $m'$ (see Section 10). The same physical conditions are formed in interstellar space. This means, first of all, that the constant speed of light from moving astrophysical objects, for example, from astrophysical jets, is a consequence of the gravidynamic effect, which fixes the relevant speed of light relative to the system of surrounding masses.

## 13 Speed of Light in a Moving Medium

Let the transparent cube move near point *B* (Figure 8). The speed of the cube is $\upsilon$ for a local observer situated at point *B*. A single photon moves inside the cube in the same direction as the cube. From time to time the photon is absorbed and after a short time is again emitted by the particles forming the cube. The conventional speed of the photon after absorption is equal to the speed of the cube. When the photon moves between absorbing particles, the local speed of the photon is rigidly fixed by the gravidynamic effect in the value *c* relative to mass *M* (see Sections 9 and 12). As a result, the average speed of the photon for a local observer becomes equal to

$$c_2 = c_1 + \upsilon(1 - c_1^2/c^2) \tag{16}$$

where $c_1$ is the average speed of the photon when the speed of the cube is zero. This result, known as the partial dragging of light by a moving medium, was observed in the water flow in the Fizeau experiment. The Fizeau optical experiment is not related to ether or to relativistic addition of speeds, but is another confirmation of the existence of the gravidynamic effect.

## 14 Mass and Energy

The gravitational acceleration of the proton and other particles with a classical inertial mass is caused by the refractive action of the gravitational field (see Section 4). This means that these particles consist of structures that, like photons, have kinetic mass and move at the speed of light. This conclusion is confirmed by some features of gravitational acceleration, which are related to the kinetic mass (see Section 5). The dilation of the proper time of these particles in gravitational fields once again confirms this conclusion, because the dilation of time is a consequence of a decrease in the local speed of light and a slowdown in local physical processes related to kinetic mass (see Sections 6 and 11). An additional confirmation of the existence of hidden structures having a kinetic mass and moving at the speed of light is an increase in the inertial mass of accelerated particles (see Section 10). All this means that only one type of mass exists in nature—kinetic mass. The classical inertial mass of particles is a consequence of combining structures having a kinetic mass (see Section 4). It follows that any potential energy, including the internal energy of the particles, is a hidden form of kinetic energy. The latter is well illustrated by the example of gravitational potential energy, behind which the kinetic energy is always hidden (see Section 7). It should be noted that kinetic energy as the only form of energy and kinetic mass as the only type of mass consistently explain the existing proportionality between the inertial mass and energy. Identical gravitational acceleration of various particles is also well explained.

## 15 Relic Radiation

The source of the cosmic microwave background radiation (CMB) at one time was moving away from us at superluminal speed; therefore the radiation going towards us at the same time should move away from us. Nevertheless, we observe this radiation because the gravidynamic effect does what it can do best—it fixes the relevant speed of photons relative to the surrounding masses (see Sections 2, 9 and 12). In intergalactic space, the role of the surrounding masses is played by the nearest galaxies and galactic groups. The gravidynamic effect fixes the relevant speed of the photons relative to the surrounding galaxies by changing the kinetic mass of photons. As a result, the speed of photons relative to the primary source of the CMB gradually increases and the kinetic mass of photons gradually decreases; but photons always move at a relevant speed in the concomitant frame of reference and ultimately reach us. During the time since the formation of the CMB, the gravidynamic effect reduced the kinetic mass of photons by more than three orders of magnitude. So, the observation of the CMB is another confirmation of the gravidynamic effect.

## 16 Heterogeneity of Time

The photon cluster very accurately simulates the temporal properties of particles with a classical inertial mass (see Section 4) because all these particles at the elementary level consist of “massless” structures that, like photons, have kinetic mass and move at the speed of light (see Section 14). If the photon cluster is located in the neighborhood of mass $M$ (Fig. 1), then the relevant speed of photons inside the cluster decreases by the gravidynamic effect to the value $c'$ (see Section 6) with some variations of this speed, depending on the distance of photons to the center of mass $M$. These local variations in the speed of light should be considered as the local heterogeneity of time in a neighborhood of mass $M$. The proper time of the photon cluster in this case is determined by the average speed of photons inside the cluster. If the photon cluster is located in the empty central area of the spherically symmetric mass $M$ (Figure 8), then the relevant speed of photons inside the cluster has a strictly constant value $c'$ (see Section 12). This allows us to speak of high homogeneity of time in this area. However, the situation changes when the photon cluster begins to move relative to mass *M*. The speed of light in this case remains constant relative to mass *M*, but not constant relative to the cluster. As a result, in the reference system associated with the photon cluster, the speed of a certain photon will depend on the direction of movement of this photon, that is, an anisotropy of the speed of light will take place. This should be considered as local anisotropy of time. In this case, the proper time of the photon cluster is also determined by the average speed of photons inside the cluster (see Section 11).

## 17 Conclusions

Time is indeed related with space, but indirectly—through movement in space. According to the concept of time considered here (see Section 6), movement at all possible structural levels causes a change in the state of the physical system, which in turn forms the proper time of this system. This concept is quite close to the relational concept of time adopted in the special theory of relativity. By the way, by the time of the creation of the general theory of relativity, one can observe a certain departure from the relational concept of time towards the substantial concept. The first step was made with the inclusion of time in a single four-dimensional space-time continuum. The last step is the appearance of gravitational waves, when space-time completely ceased to depend on physical systems. Within the framework of the new model of gravitation and the new concept of time, the gravitational wave is a wave consisting of two gravitational effects—gravicontraction and gravidynamic. The first effect acts on the space. The second effect acts on the speed of light, which in turn effects the speed of the time flow. Now our task is to reliably prove the existence of the gravidynamic effect.

The existence of the gravidynamic effect is confirmed by a very large number of experimental facts (see Sections 2-15). One confirmation of this effect is an increase in the inertial mass of a test particle accelerated relative to the surrounding masses (see Section 10). The dilation of the proper time of the test particle also confirms the existence of the gravidynamic effect (see Section 6); moreover, double confirmation takes place (see Section 11). The constant speed of light observed in different conditions is another confirmation of the gravidynamic effect (see Sections 9 and 12). The partial entrainment of light by the moving medium also does not remain aside (see Section 13). Even cosmic microwave background radiation comes to us due to the participation of this effect (see Section 14). However, the gravidynamic effect does not always work alone. For example, a remotely observed decrease in the speed of light in gravitational fields is formed by two gravitational effects—gravidynamic and gravicontraction (see Section 2). In addition, these two primary gravitational effects create a wide range of secondary effects, including gravitational refraction (see Section 3) and gravitational acceleration (see Section 4).

An additional confirmation of the gravidynamic effect is the Sagnac experiment, which demonstrates the constancy of the local speed of light relative to the Earth and non-constancy relative to the measuring device. The same result, namely, anisotropy of the speed of light, can be observed in the reference system related with a proton accelerated at the Large Hadron Collider. In this case, anisotropy is also caused by the constancy of the local speed of light relative to the Earth (see Section 9). Thus, the decisive confirmation of the gravidynamic effect can be realized, not near the neutron star but on Earth. The experiment can be performed in the framework of the testing of the special theory of relativity. The anisotropy of the speed of light is not the only surprise produced by gravitation. For example, in the reference system related with the same proton accelerated at the LHC, the local time in Geneva, contrary to expectations, will be observed accelerated in the same proportion as the dilation of the proper time of the proton. In addition, any change in the proper time of the oncoming proton will not be observed at all, despite the very high relative speed. This is due to the fact that the dilation of the proper time of protons is a consequence of the gravidynamic effect (see Section 11) and both counter protons are in strictly identical conditions regarding this effect.